\documentclass[aps,pra,twocolumn,groupedaddress,showpacs]{revtex4}

\usepackage{graphicx}
\usepackage{amsmath}
\usepackage{bm}

\newcommand{\be}{\begin{equation}}
\newcommand{\ee}{\end{equation}}
\newcommand{\bwt}{\begin{widetext}}
\newcommand{\ewt}{\end{widetext}}
\newcommand{\bea}{\begin{eqnarray}}
\newcommand{\eea}{\end{eqnarray}}

\newcommand{\eqn}[1]{\begin{equation}#1 \end{equation}}

\begin{document}
\pagenumbering{arabic}

\title{Perpendicular laser cooling with a rotating wall potential in a Penning trap}

\author{Steven B. Torrisi$^{1,2}$}
\author{Joseph W. Britton$^{1,3}$}
\author{Justin G. Bohnet$^{1}$}
\author{John J.~Bollinger$^{1}$}
\affiliation{\mbox{$^1$National Institute of Standards and Technology, Boulder, CO 80305, USA}\\
\mbox{$^2$Department of Physics and Astronomy, University of Rochester, Rochester, NY, 14627, USA }\\
\mbox{$^3$Army Research Lab, Adelphi, MD 20783}}

\begin{abstract}
We investigate the impact of a rotating wall potential on perpendicular laser cooling in a Penning ion trap. By including energy exchange with the rotating wall, we extend previous Doppler laser cooling theory and show that low perpendicular temperatures are more readily achieved with a rotating wall than without. Detailed numerical studies determine optimal operating parameters for producing low temperature, stable 2-dimensional crystals, important for quantum information processing experiments employing Penning traps.

\end{abstract}


\maketitle

\section{Introduction}
Ion crystals confined in Penning traps provide opportunities for
interesting studies in atomic physics \cite{tanj95, grul01,shin11,andz13}, quantum information
\cite{biem09b,brij12,wanc13,mavs13,bohj16}, and plasma physics \cite{andf09,jenm05,sawb12}.
Doppler laser cooling is the principle cooling technique used to generate
the crystals. When the thermal energy of the ions is small compared
to their Coulomb potential energy, ion crystals naturally form in order
to minimize the ion Coulomb potential energy. Stable, low-temperature crystals are crucial
for some applications. For example, simulations of quantum magnetism require stable single-plane crystals for single-ion detection. Low ion temperatures improve both the crystal stability and the fidelity of the simulations \cite{brij12,sawb14,bohj16}.

Doppler laser cooling in a Penning trap is complicated by
the fact that the ion crystals rotate, producing large, coherent Doppler
shifts \cite{itaw82,itaw88,hen08,aspm14}.
A theoretical treatment of this complication for a many-ion crystal (or a cold non-neutral plasma)
and the resulting minimum attainable temperatures were discussed
more than two decades ago \cite{itaw88}. Here we update this theory,
taking into account an important experimental advance.
Specifically, current experiments routinely apply sinusoidal potentials to azimuthally
segmented trap electrodes to generate a rotating potential \cite{huap97,huap98b,mitt98,bhas12}.
This rotating potential, frequently called a rotating wall, sets the rotation frequency and applies a torque
that balances the torque imparted by the perpendicular laser cooling
beam. We update previous
theory \cite{itaw88} by accounting for the work done by the rotating wall in applying
a torque, and investigate in some detail the conditions that minimize the temperature.
In addition we investigate the dependence of the Doppler cooling laser-beam
torque on laser parameters and crystal rotation frequency. Conditions
for minimal torque minimize shear stress, and may help in achieving
stable crystals.

\begin{figure}
\includegraphics[scale=1.0]{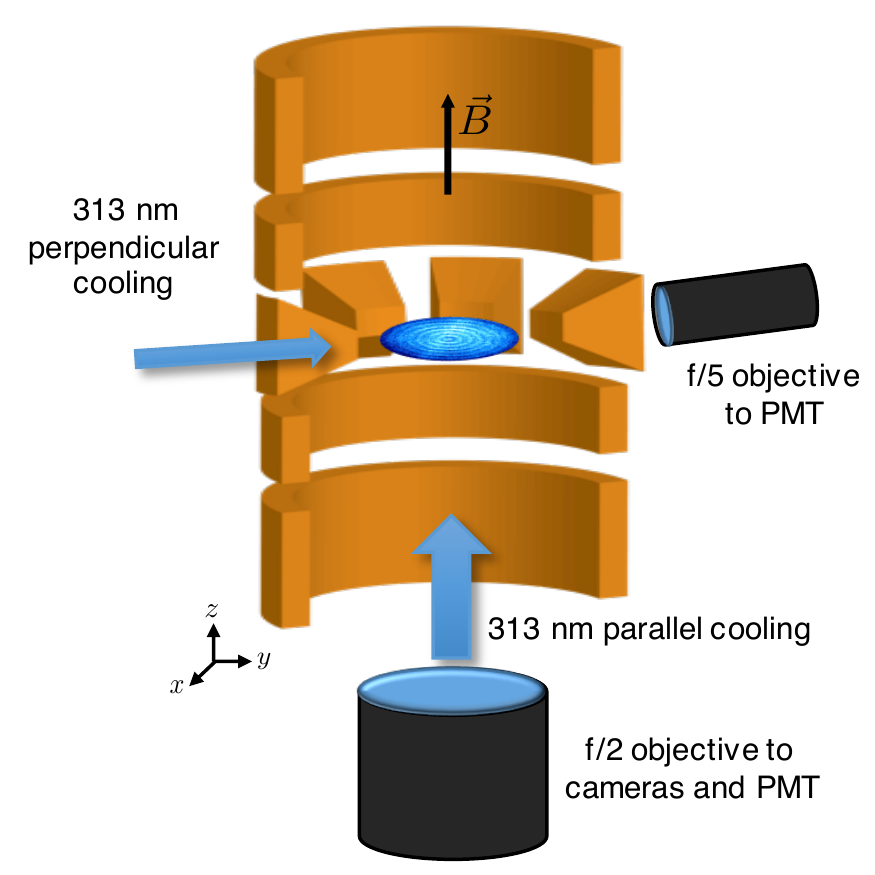}
\caption{Cross sectional sketch of the NIST Penning trap used to generate and
control single-plane crystals. Electrostatic potentials applied to cylindrical electrodes generate
a confining well in the direction of the trap symmetry ($\hat{z}$) axis. The trap is immersed in a
strong uniform magnetic field ($B\approx4.5$ T) directed parallel to the $\hat{z}$-axis.
Radial confinement is due to the Lorentz force generated by $\vec{E}\times\vec{B}$-induced
rotation through the magnetic field. The central ``ring'' electrode
is segmented into eight sections, and used to apply a rotating wall potential.
Doppler laser cooling beams are directed both along the trap axis
and perpendicular to the trap axis. \label{fig:Cross-sectional-sketch}}

\end{figure}

We study the single-plane crystal geometry sketched
in Fig.\ \ref{fig:Cross-sectional-sketch} that is used in current
National Institute of Standards and Technology (NIST) quantum simulation experiments \cite{brij12,sawb14,bohj16}. In this
work, Doppler laser cooling is provided by laser beams directed parallel
and perpendicular to the magnetic field, which is oriented in the $\hat{z}$ direction.
We neglect any coupling between
the in-plane (radial) and out-of-plane (axial) degrees of freedom,
which should be weak for this single-plane geometry. We also initially
neglect recoil heating of the in-plane degrees of freedom by the
parallel laser beam. In this case the temperature of the in-plane
degrees of freedom will be determined by the perpendicular laser and
the problem is reduced to the 2-dimensional geometry diagrammed in Fig.
2. After analyzing this problem we add in the effects of scattering
recoil from the parallel Doppler cooling laser beam.

\begin{figure}
\includegraphics[scale=0.45]{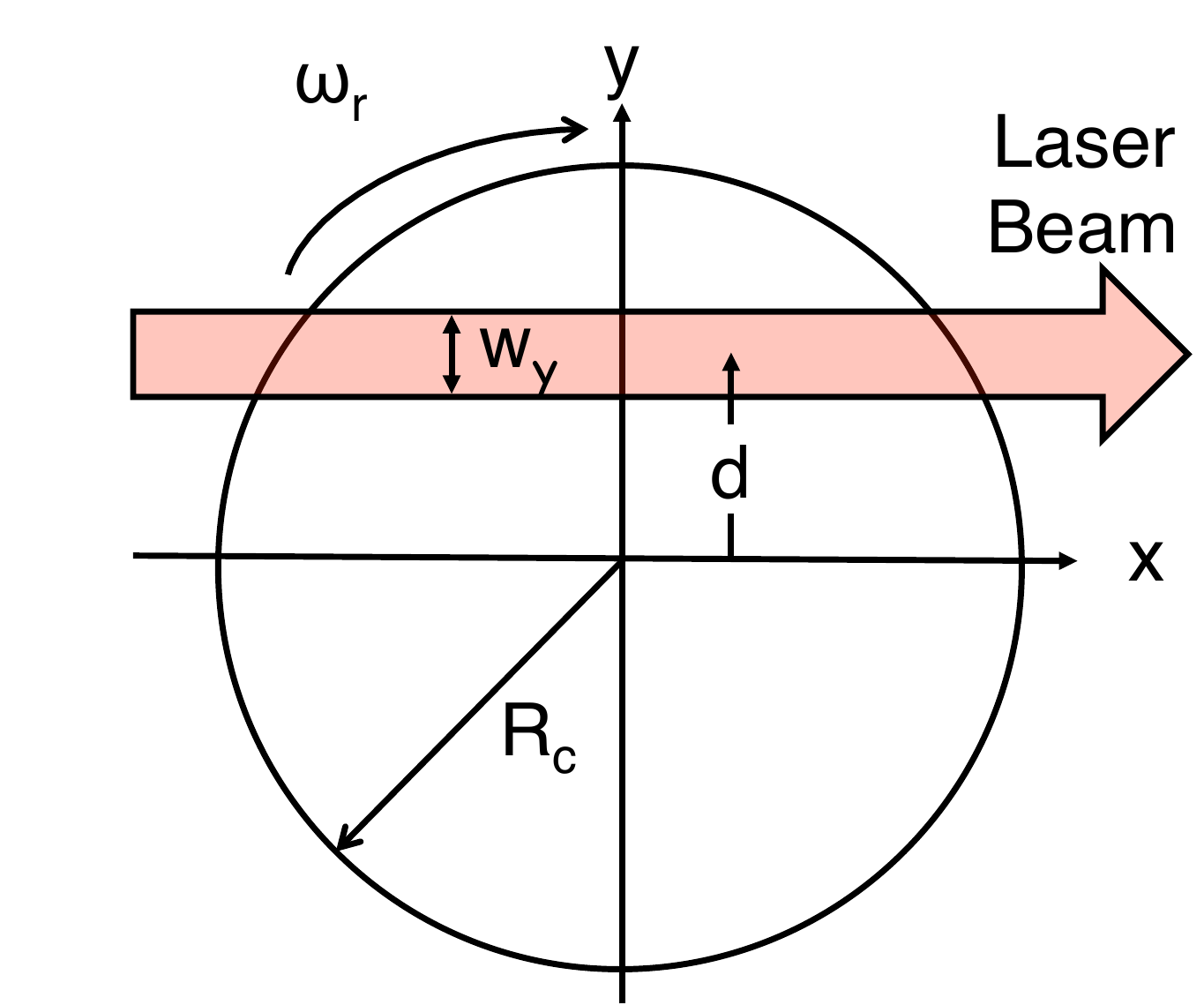}\caption{2-dimensional geometry for perpendicular laser cooling of a single plane array of radius $R_c$. The array rotates at the angular frequency $\omega_r$.
The cooling laser is directed parallel to $\hat{x}$ with wave vector $\vec{k}=k\hat{x}$.
$d$ denotes the offset of the perpendicular Doppler cooling
laser beam from the center of the array. For $d > 0$ the laser beam
is directed to the side of the array that is receding from the laser
beam due to the array rotation. A Gaussian beam profile $I(y)=I_{0}\exp[\frac{-2(y-d)^2}{{w_y}^2}]$ with waist $w_y$ is assumed. For the Doppler cooling limit calculation
the array is treated as a continuous medium with
a 2-d density $\sigma(x,y)$ as discussed in the text. \label{fig:2d-geometry}}

\end{figure}

A main conclusion of this manuscript is that low in-plane temperatures
are more readily achieved with a rotating wall than without. Itano
\cite{itaw88} provided a careful theoretical and experimental study
of perpendicular laser cooling in a Penning trap without a rotating
wall and noted that, for a wide range of operating parameters, the
minimum perpendicular temperature was two orders of magnitude larger
than the normal single-ion Doppler cooling limit $T_{\mathrm{Doppler}}=\hbar\gamma_{0}/2k_{B}$
($\approx0.44$ mK for the $2s^{2}S_\frac{1}{2} \rightarrow 2p^{2}P_\frac {3}{2}$ laser cooling transition in $^{9}$Be$^{+}$). These higher temperatures are due to work done by the perpendicular laser in applying a torque that is necessary,
in equilibrium, to balance ambient torques due to static field errors and asymmetries.

In general, static field errors
due to imperfect trap construction result in a torque that tends to slow the crystal rotation \cite{dubd99}. Without
a rotating wall, equilibrium, which is characterized by a constant rotation frequency $\omega_r$ and
constant crystal radius $R_c$, occurs when the perpendicular laser beam applies a
counterbalancing torque $\tau_{\mathrm{Laser}}$ that has the same sign
as the rotation $\omega_{r}$. In applying this torque the laser beam performs work,
increasing the energy of the crystal. For constant $\omega_{r}$ and $R_c$
this energy input must go into increasing the in-plane thermal energy
of the ions.  Although static field errors generate a torque, they can only convert ion potential energy to ion thermal energy, or vice versa \cite{dubd99}. Therefore, in equilibrium static field errors can not change the ion thermal energy. In contrast, the rotating wall
is generated with time-dependent potentials and can perform work, entering
into the overall energy balance. In equilibrium the torque due to
the rotating wall opposes the
laser beam torque. The rotating wall can therefore provide an energy sink for the work done
by the laser beam torque, resulting in lower in-plane temperatures.

In Sec.\ \ref{sec:simpletheory}, we write down expressions for the perpendicular laser
scatter rate and torque, and discuss a simple theory for the energy
balance between perpendicular Doppler laser cooling and the work done
by the rotating wall. We also
discuss recoil heating from the parallel laser beam. In Sec.\ \ref{sec:numerics},
we present numerical studies that vary the perpendicular
cooling laser beam offset and waist, as well as the rotation rate of the crystal. These
studies suggest optimal operating parameters that result in low in-plane
temperatures and low laser beam torques. In Sec.\ \ref{sec:conclusion} we present
some concluding remarks.

\section{Simple Theory \label{sec:simpletheory}}

In this section we adapt the formalism of Itano \cite{itaw88} to the experimental context of a 2-dimensional Coulomb crystal and obtain expressions for the photon scatter rate, rate of energy exchange, and torque due to a perpendicular cooling laser. We then add in an energy exchange contribution due to the rotating wall, and finally a recoil heating term due to a parallel cooling laser beam.  Although the perpendicular cooling laser and rotating wall provide localized heating or cooling, we assume the in-plane degrees of freedom equilibrate rapidly and can be characterized by thermal equilibrium at a temperature $T_{\perp}$ \cite{footnote.thermalCOM}. We are interested in determining $T_{\perp}$ as a function of the parameters outlined in Fig.\ \ref{fig:2d-geometry}.  We neglect any coupling between the in-plane degrees of freedom and the parallel (to the $\hat{z}$-axis) degrees of freedom, anticipated to be a good approximation for single plane crystals.

\subsection{Scattering Rate}

The expression for the photon scattering rate $\gamma_L$ for an ion with mass $m$, position $(x,y)$, and velocity $(v_x,v_y)$ in the laser beam of Fig. \ref{fig:2d-geometry} is \cite{itaw88}

\eqn{\gamma_L(y,v_x) = \frac{I(y) \sigma_0}{\hbar \omega_L} \frac{\left(\gamma_0/2 \right)^2}{(\gamma(y)/2)^2 + (\Delta\omega-k v_x)^2}\, , \label{eq:scatter1}}
where $I(y)$ is
the laser intensity, $\sigma_0$ is the scattering cross-section on resonance, $\omega_L$ is the angular frequency
of the laser light, $\gamma_0$ is the natural linewidth of the targeted transition, and $\gamma(y)$ is the linewidth adjusted for saturation. $\Delta\omega=\omega_L-\omega_0-R/\hbar$ is the detuning of the laser from the atomic transition frequency $\omega_0$, taking recoil $R=(\hbar k)^2/2m$ into account. $k$ is the wave vector of the laser light, which we approximate as $k = \omega_0/c$, the wave vector on resonance, throughout the manuscript. We assume the laser beam has a Gaussian beam profile $I(y)=I_{0}\exp[\frac{-2(y-d)^2}{{w_y}^2}]$ and is directed parallel to $\hat{x}$, in which case $kv_x$ is the Doppler shift. The saturation adjusted linewidth can be written in terms of the saturation parameter $S(y)=I(y)\sigma_0/(\hbar\omega_0\gamma_0)$ as $\gamma(y)^2={\gamma_0}^2(1+2S(y))$. With $S_0\equiv I_0\sigma_0/(\hbar\omega_0\gamma_0)$ denoting the saturation parameter at maximum laser beam intensity, Eq. (\ref{eq:scatter1}) can be rewritten as
\eqn{\gamma_L=\frac{\omega_0}{\omega_L}\frac{\gamma_0 S_0\exp[\frac{-2(y-d)^2}{{w_y}^2}]}{1+2S_0  \exp[\frac{-2(y-d)^2}{{w_y}^2}]+(\frac{2}{\gamma_0})^2(\Delta\omega-k v_x)^2}\, . \label{eq:scatter2}}
In the remainder of the manuscript we neglect the $\omega_0/\omega_L$ multiplicative factor as it is very close to unity for the detunings we consider.

For calculating the total torque and energy exchange with the perpendicular laser beam we approximate the ion crystal as a continuous medium with areal density
\eqn{\sigma(x,y)= \Sigma_0 \sqrt{1-\frac{x^2+y^2}{R_c^2}}\, , \label{eq:areal density}}
where $\Sigma_0$ is the areal density at the center of the crystal. Equation (\ref{eq:areal density}) assumes a quadratic trap potential and is valid for a single-plane non-neutral plasma with sufficiently low temperature that the Debye length is small compared to $R_c$. It results from the projection of a uniform density spheroid onto $z=0$ \cite{dubd99,dubd13}.

The scattering rate $Sc(x,y)$ per unit area is obtained by averaging the product of Eqs. (\ref{eq:scatter2}) and (\ref{eq:areal density}) with a thermal distribution of velocities $v_x$. Thermal equilibrium is characterized by a Maxwell-Boltzmann velocity distribution super-imposed on rigid-body rotation at $\omega_r$ \cite{dubd99}. For the $\hat{x}$-component of the velocity, this reduces to
\eqn{P(v_x | y, u) = \exp{ \left[\frac{ - (v_x - \omega_r y)^2}{u^2} \right] }/  (u\sqrt{\pi})\, ,}
where
\eqn{u\equiv\sqrt{2 k_{\mathrm{B}}T_{\perp}/m}\, .\label{eq:mean_velocity}}
The expression
\eqn{Sc(x,y)=\int_{-\infty}^{\infty} \mathrm{d}v_x P(v_x | y, u) \sigma(x,y) \gamma_L(y,v_x)}
can be integrated over the crystal to obtain the total photon scattering rate by the crystal.

\subsection{Torque and Energy Transfer}
We are particularly interested in expressions for the torque and energy exchange rate produced by the perpendicular laser cooling beam. The change in an ion's momentum per photon scattering event, averaged over many scattering events, is $\left< \Delta \vec{p} \right> =\hbar k \hat{x}$. If these scattering events occur at the position $(x,y)$ in the crystal, they impart an average angular momentum per scattering event of $\hbar k y$. This transfer of angular momentum results in a torque $\tau_{\mathrm{Laser}}$ imparted by the laser beam on the crystal given by
\eqn{\tau_{\mathrm{Laser}} = \int_{-R}^{R} \mathrm{d}x \int_{-\sqrt{R_c^2-x^2}}^{\sqrt{R_c^2-x^2}}  \mathrm{d}y\, \hbar k y\, Sc(x,y)\, . \label{eqn:torque}}
Note that we implicitly use the convention that $\omega_r > 0$, in which case $\tau_{\mathrm{Laser}} > 0$ when the laser applies a torque that tends to increase the rotation frequency, obtained for $d \gg 0$ like that shown in Fig.\ \ref{fig:2d-geometry}.

Let $\left\langle\Delta E_{K,\perp} \right\rangle$ denote the average change in the in-plane kinetic energy of an ion for a photon scattering event in the perpendicular laser beam. We determine the rate of energy change of the ion crystal produced by the perpendicular laser by multiplying $\left\langle\Delta E_{K,\perp} \right\rangle$ with $P(v_x | y, u)$, $\sigma(x,y)$, and $\gamma_L(y,v_x)$ and integrating over velocities and spatial coordinates. In three dimensions, $\left\langle\Delta E_K\right\rangle=\hbar \vec{k}\cdot\vec{v}+2R$ \cite{itaw88}, where the recoil heating term $2R$ has equal contributions from photon absorption and emission. For simplicity we assume an isotropic distribution of the scattered photons, in which case on average $\frac{2}{3}$ of the emitted photon recoil gets shared with the in-plane degrees of freedom, and $\frac{1}{3}$ gets shared with the axial (or transverse) degrees of freedom \cite{itaw82}.  Because we assume no coupling between the in-plane and transverse degrees of freedom, we have $\left\langle\Delta E_{K,\perp} \right\rangle = \hbar k v_x + \frac{5R}{3}$, resulting in a total rate of energy change $\left\langle\frac{\mathrm{d}E}{\mathrm{d}t}\right\rangle_{\mathrm{Laser}}$ due to the cooling laser of
\begin{multline}
\left\langle\frac{\mathrm{d}E}{\mathrm{d}t}\right\rangle_{\mathrm{Laser}} = \int_{-R}^{R} \mathrm{d}x \int_{-\sqrt{R_c^2-x^2}}^{\sqrt{R_c^2-x^2}}  \mathrm{d}y \int_{-\infty}^{\infty} \mathrm{d}v_x\
\\
\\
\times \left(\hbar k v_x + \frac{5R}{3} \right) P(v_x | y, u) \sigma(x,y) \gamma_L (y, v_x)\, . \label{eqn:laser rate}
\end{multline}


\subsection{Energy Exchange with a Rotating Wall}
The rotating wall performs work in applying a torque that maintains a constant rotation frequency. Experimental observation of stable rotation frequencies implies that the net torque on the crystal must be balanced. We assume the rotating wall and the laser provide the largest contributions to the torque, and neglect other sources of torque, such as gas collisions or static field asymmetries. This assumption is supported for our set-up by the observation of very slow evolution (time scale $\sim$ minutes) of the ion cloud equilibrium in the absence of the perpendicular cooling laser and the rotating wall \cite{jenm04}. Therefore, in equilibrium $\tau_{\mathrm{Wall}} = -\tau_{\mathrm{Laser}}$, and the work done by the rotating wall is
\eqn{\left\langle \frac{\mathrm{d}E}{\mathrm{d}t} \right\rangle_{\mathrm{Wall}} = - \omega_{r} \tau_{\mathrm{Laser}}\, , \label{eqn:torquetimesomega}}
where the torque of the laser is obtained from Eq.\ (\ref{eqn:torque}). Note that for $\tau_{\mathrm{Laser}} > 0$ (configuration of Fig.\ \ref{fig:2d-geometry}) the rotating wall is an energy sink, producing lower $T_{\perp}$.

Equations (\ref{eqn:torque}), (\ref{eqn:laser rate}), and (\ref{eqn:torquetimesomega}) can be combined to give an expression for the total energy balance due to both the laser and the rotating wall,
\begin{widetext}
\begin{multline}
\left\langle\frac{\mathrm{d}E}{\mathrm{d}t}\right\rangle_{\mathrm{Laser-Wall}} = \left\langle\frac{\mathrm{d}E}{\mathrm{d}t}\right\rangle_{\mathrm{Laser}} + \left\langle\frac{\mathrm{d}E}{\mathrm{d}t}\right\rangle_{\mathrm{Wall}} =
\\
\\
\int_{-R_c}^{R_c} \mathrm{d}x \int_{-\sqrt{R_c^2-x^2}}^{\sqrt{R_c^2-x^2}}  \mathrm{d}y \ \int_{-\infty}^{\infty} \mathrm{d}v_x\ \left(\hbar k \left(v_x- \omega_r y\right) + \frac{5R}{3} \right)
P(v_x | y, u) \sigma(x,y) \gamma_L (y, v_x) \, .
\label{eqn:total energy rate}
\end{multline}
\end{widetext}


For numerical calculation we make the substitution $v = (v_x-\omega_r y)/u$, which simplifies the Maxwell-Boltzmann distribution and the term for the energy change per scattering event, while adjusting the denominator of $\gamma_L (y, v_x)$. Substituting for $R = (\hbar k)^2/(2m)$, we obtain,
\begin{widetext}
\begin{equation}
\left<\frac{\mathrm{d}E}{\mathrm{d}t} \right>_{\mathrm{Laser-Wall}} =
\int_{-R_c}^{R_c} \mathrm{d}x \int_{-\sqrt{R_c^2-x^2}}^{\sqrt{R_c^2-x^2}}  \mathrm{d}y \int_{-\infty}^{\infty} \mathrm{d}v \frac{\exp [-v^2 ]}{\sqrt{\pi}}
\frac{ (v+ \frac{5 \hbar k }{6 m u}) \gamma_0 S_0 \Sigma_0
 \exp [ \frac{-2(y-d)^2}{ w_y^2}] \sqrt{1-\frac{x^2+y^2}{R_c^2}}
 }
 { \left[ 1+2S_0 \exp [  \frac{-2(y-d)^2}{ w_y^2}] +\left( \frac{2}{\gamma_0} \right)^2  (\Delta \omega - k (\omega_r y +v u))^2 \right]}. \label{eqn:finalenergy}
\end{equation}
\end{widetext}
Roots of $\left<\frac{\mathrm{d}E}{\mathrm{d}t} \right>_{\mathrm{Laser-Wall}} = 0$ as a function of $u$ determine the Doppler cooling limit through $T_{\perp} = \frac{u^2 m}{2k_B}$ (Eq.\ \ref{eq:mean_velocity}).

\subsection{Recoil Heating from a Parallel Cooling Beam \label{sec:recoil}}
Our primary interest is in the roots of Eq.\ (\ref{eqn:finalenergy}). However, our Penning trap features a parallel cooling laser beam (see Fig. \ref{fig:Cross-sectional-sketch}) that Doppler cools ion motion parallel to $\hat{z}$ (the magnetic field direction) but, through photon recoil, heats the in-plane motion. We approximately model this recoil heating to determine under what conditions it can significantly elevate the in-plane temperature.

We account for recoil heating with a parallel cooling laser by adding a constant term $\left<\frac{\mathrm{d}E}{\mathrm{d}t} \right>_{\parallel}$ to the energy balance equation (Eq.\ (\ref{eqn:finalenergy})). We estimate this term with some simplifying assumptions. First, we assume the parallel laser beam waist is large compared to $R_c$ (typical for our set-up), so that the parallel laser beam intensity $I_{\parallel}$  and saturation parameter $S_{\parallel}=I_{\parallel} \sigma_0/(\hbar \omega_{\parallel} \gamma_0)$ can be treated as constants for all ions in the crystal. We assume the parallel laser beam frequency $\omega_{\parallel}$ is, in general, different from the perpendicular laser beam frequency, and set $\omega_{\parallel}$ half a linewidth below the atomic transition frequency, $\omega_{\parallel}-\omega_0 = -\gamma_0/2$, where the Doppler laser cooling limit for $T_{\parallel}$ is achieved \cite{sawb12,sawb14,mavs14}.  At the Doppler cooling limit the thermal broadening of the natural Lorentzian line profile is small.  We therefore neglect Doppler shifts and obtain a parallel laser scatter rate per ion of
\eqn{\gamma_{L,\parallel} = \frac {\gamma_0 S_{\parallel}}{2+2S_{\parallel}}\, .}
With the assumption of isotropic scattering, on average $\frac{2}{3}$ of the emitted photon recoil gets shared with the
in-plane degrees of freedom. Finally $\gamma_{L,\parallel}$ gets multiplied by the number of ions $N=\Sigma_0 \frac {2}{3} \pi {R_c}^2$ in the crystal, obtained by integrating Eq. (\ref{eq:areal density}).  This provides the following estimate for the parallel laser recoil heating rate,
\eqn{\left<\frac{\mathrm{d}E}{\mathrm{d}t} \right>_{\parallel} = \frac {\gamma_0 S_{\parallel}}{1+S_{\parallel}} \frac {R}{3} \left(\Sigma_0 \frac{2}{3} \pi {R_c}^2\right)\, .}
In our approximate treatment of recoil heating from the parallel cooling beam we have ignored the additional saturation of the cooling transition for ions that are interacting with both laser beams at the same time. Our treatment should be reasonably accurate for low parallel laser intensities.

\subsection{Simplifications for $|d|, w_y \ll R_c$ }
Equation (\ref{eqn:finalenergy}) can be simplified in the limit $|d|, w_y \ll R_c$, conditions typically employed in experiments.  This enables a more transparent determination of the dependence of the Doppler cooling limits on combinations of parameters.  For $|d|, w_y \ll R_c$, contributions to the integral in Eq.\ (\ref{eqn:finalenergy}) occur predominantly for $|y| \ll R_c$. Over this range we separate the 2-d density $\sigma (x,y)$ into a product of terms that separately depend on $x$ and $y$,
\eqn{\sqrt{1- \frac{x^2+y^2}{R_c^2}} \approx \sqrt{1- \frac{x^2}{R_c^2}} \sqrt{1- \frac{y^2}{R_c^2}}\, .}
 The region where this approximation fails $(|x|\sim R_c)$ is small, and also characterized by low areal density, therefore not significantly contributing to the integral. The $\sqrt{1- \frac{x^2}{R_c^2}}$ term can be factored outside the $y$ and $v$ integrals, and with the assumption $|d|, w_y \ll R_c$, the limits of the $y$ integral can be extended to $(-\infty, +\infty)$,
\begin{widetext}
\begin{multline}
\left<\frac{\mathrm{d}E}{\mathrm{d}t} \right>_{\mathrm{Laser-Wall}} \approx \frac{\gamma_0 S_0 \Sigma_0}{\sqrt{\pi}} \int_{-R}^{R} \mathrm{d}x \sqrt{1- \frac{x^2}{R_c^2}}
\\
\times \int_{-\infty}^{\infty}  \mathrm{d}y \int_{-\infty}^{\infty} \mathrm{d}v \sqrt{1- \frac{y^2}{R_c^2}} \exp [ \frac{-2(y-d)^2}{ w_y^2}]
\frac{ (v+ \frac{5 \hbar k }{6 m u}) \exp [-v^2 ]}
 { \left[ 1+2S_0 \exp [  \frac{-2(y-d)^2}{ w_y^2}] +\left( \frac{2}{\gamma_0} \right)^2  (\Delta \omega - k (\omega_r y +v u))^2 \right]}. \label{eqn:simplification 1}
\end{multline}
\end{widetext}


The bottom line of Eq.\ (\ref{eqn:simplification 1}) no longer depends on $x$.  It is clear that roots to Eq.\ (\ref{eqn:simplification 1}) must be roots of the bottom line involving integrals only over $y$ and $v$.  For now we approximate $\sqrt{1-y^2/R_c^2} \approx 1$ and write the bottom line in terms of dimensionless parameters and variables,
\begin{widetext}
\begin{equation}
\int_{-\infty}^{\infty}  \mathrm{d}\delta \int_{-\infty}^{\infty} \mathrm{d}v \: \mathrm{e}^{-2\delta^2} \mathrm{e}^{-v^2}
   \frac{ (v+ v_{\mathrm{rec}}/u)}
   { \left[ 1+2S_0 \mathrm{e}^{-2\delta^2} + (\frac{\Delta \omega -k \omega_r d}{\gamma_0/2} - \frac{k \omega_r w_y}{\gamma_0/2}\delta - \frac{k v_{\mathrm{rec}}}{\gamma_0/2}\frac{u}{v_{\mathrm{rec}}} v)^2 \right]}. \label{eqn:dimensionless}
\end{equation}
\end{widetext}
Here $v_{\mathrm{rec}}\equiv 5\hbar k/(6m)$ is an effective recoil velocity, $\delta\equiv(y-d)/w_y$, and we neglect any multiplicative constants. Roots of Eq.\ (\ref{eqn:dimensionless}) for $u/v_{\mathrm{rec}}$ depend on the single-ion parameters $S_0$ and $(k v_{\mathrm{rec}})/(\gamma_0/2)$, and on two parameters that depend on properties of the crystal and the laser beam.  They are the detuning of the laser from the Doppler-shifted atomic resonance at the center of the laser beam,
\eqn{\Delta_d\equiv\frac{\Delta \omega -k \omega_r d}{\gamma_0/2}\, , \label{eqn:detuning}}
 and the dispersion in the Doppler shift across the laser beam waist,
\eqn{\Delta_{w}\equiv\frac{k\omega_r w_y}{\gamma_0/2}\, . \label{eqn:dispersion}}
Both of these parameters are normalized to the half linewidth $\gamma_0/2$ of the atomic transition.
The equilibrium temperatures predicted by Eq. (\ref{eqn:dimensionless}) are independent of the crystal radius $R_c$.

We can include the lowest order correction to the dependence of the density on $y$, $\sqrt{1-y^2/R_c^2} \approx 1-\frac{1}{2}\frac{y^2}{R_c^2} \approx \exp(-\frac{1}{2}\frac{y^2}{R_c^2})$. This term can be combined with the multiplicative $\exp\left[\frac{-2(y-d)^2}{w_y^2}\right]$ term in Eq.\ (\ref{eqn:simplification 1}). After adding the exponents, expanding, and completing the square, we obtain, up to a constant multiplicative factor, another Gaussian beam profile $\exp\left[\frac{-2(y-d')^2}{{w'}_y^2}\right]$ with a rescaled position $d'$ and waist ${w'}_y$,
\begin{align}
   {w'}_y^2 &=\frac{1}{1+\frac{{w_y}^2}{4{R_c}^2}} w_y^2 \label{eqn:wy prime}\\
   d' &= \frac{1}{1+\frac{{w_y}^2}{4{R_c}^2}} d \, . \label{eqn:d prime}
\end{align}
If the saturation parameter $S_0$ is sufficiently small so that the $2S_0 \exp [  \frac{-2(y-d)^2}{ w_y^2}]$ term in the denominator of Eq.\ (\ref{eqn:simplification 1}) can be neglected, then the Doppler laser cooling limits will depend on the same combination of parameters as Eqs.\ (\ref{eqn:detuning}) and (\ref{eqn:dispersion}) but with $d$ and $w_y$ replaced by $d'$ and $w'_y$.

\section{Numerical Studies \label{sec:numerics}}
We use the theory of the previous section to conduct a numerical study of the Doppler laser cooling limit and laser beam torque as a function of experimental parameters. We start by determining the Doppler laser cooling limits through the roots of the Eq.\ (\ref{eqn:dimensionless}), valid for $|d|, w_y \ll R_c$. We then present some examples of Doppler laser cooling limits obtained with the more general Eq.\ (\ref{eqn:finalenergy}), and assess the region of validity of the simplifying $|d|, w_y \ll R_c$ assumption. Calculations of the laser torque from  Eq.\ (\ref{eqn:torque}) are used to determine conditions that minimize shear stress on the crystal.

\subsection{Doppler cooling limits for $|d|, w_y \ll R_c$}

\begin{figure}[htb]
\includegraphics[scale=0.33]{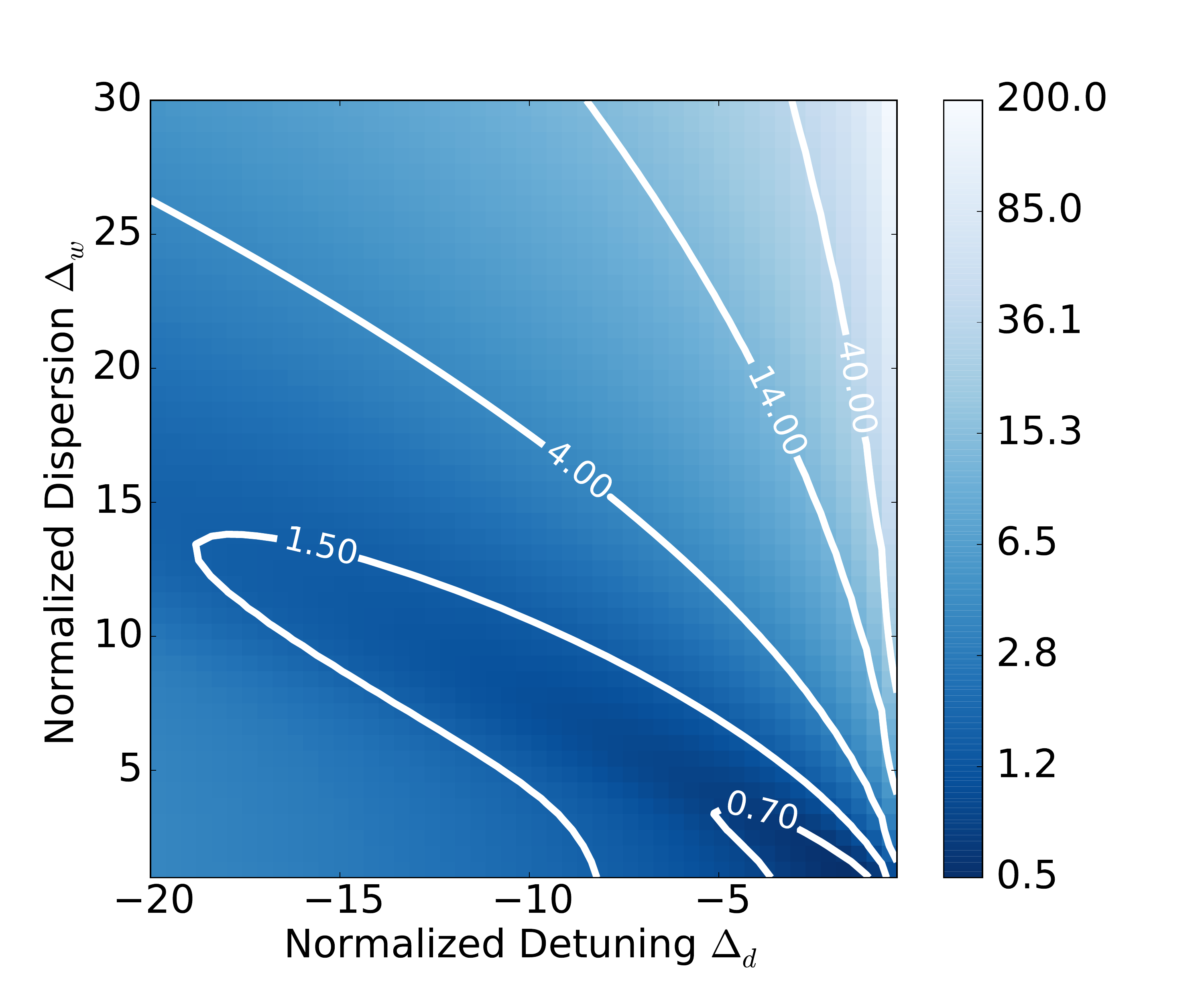}
\caption{Equilibrium temperature $T_{\perp}$ as a function of the normalized detuning $\Delta_d$ (Eq.\ (\ref{eqn:detuning})) of the laser from the Doppler-shifted atomic resonance and the normalized dispersion $\Delta_{w}$ (Eq.\ (\ref{eqn:dispersion})) in the Doppler shift across the laser beam waist. The units of the labeled color bar and contours are millikelvin. The calculation uses Eq.\ (\ref{eqn:dimensionless}), valid for $|d|, w_y \ll R_c$, to determine the equilibrium temperature, and assumes $S_0=0.5$ and parameters for $^{9}$Be$^{+}$ . The single-ion Doppler laser cooling limit for $^{9}$Be$^{+}$ is 0.44 mK.
\label{fig:small-d-w}}
\end{figure}
Figure \ref{fig:small-d-w} displays the Doppler laser cooling limits obtained from the values of $u/v_{\mathrm{rec}}$ that cause expression (\ref{eqn:dimensionless}) to vanish. We use $^{9}$Be$^{+}$ as an example, where, for the $2s^{2}S_\frac {1}{2} \rightarrow 2p^{2}P_\frac{3}{2}$ laser cooling transition, $\lambda=2\pi/k= 313$ nm, $\gamma_0/(2\pi) = 18$ MHz, and $v_{\mathrm{rec}} = 0.118$ m/s. For a given $\Delta_{w}$, the equilibrium temperature $T_{\perp}$ goes through a minimum at a value of $\Delta_d$ that is slightly larger than $\Delta_{w}$. This minimum grows with increasing $\Delta_{w}$. Low $T_{\perp}$ within a factor of two of the 0.44 mK single-ion Doppler laser cooling limit is obtained for a range of values for $\Delta_d$ and $\Delta_{w}$ satisfying $-5<\Delta_d<-1$ and $\Delta_{w}< 5$.  For $\Delta_{w} \gg \Delta_d$ (upper right hand corner of Fig.\ \ref{fig:small-d-w}) large values of $T_{\perp}$ are obtained, as this condition results in some ions scattering laser light that is blue detuned to the Doppler shifted atomic transition frequency.

\subsection{General cooling limits and applied laser torque}

\begin{figure}[htb]
\includegraphics[scale=0.33]{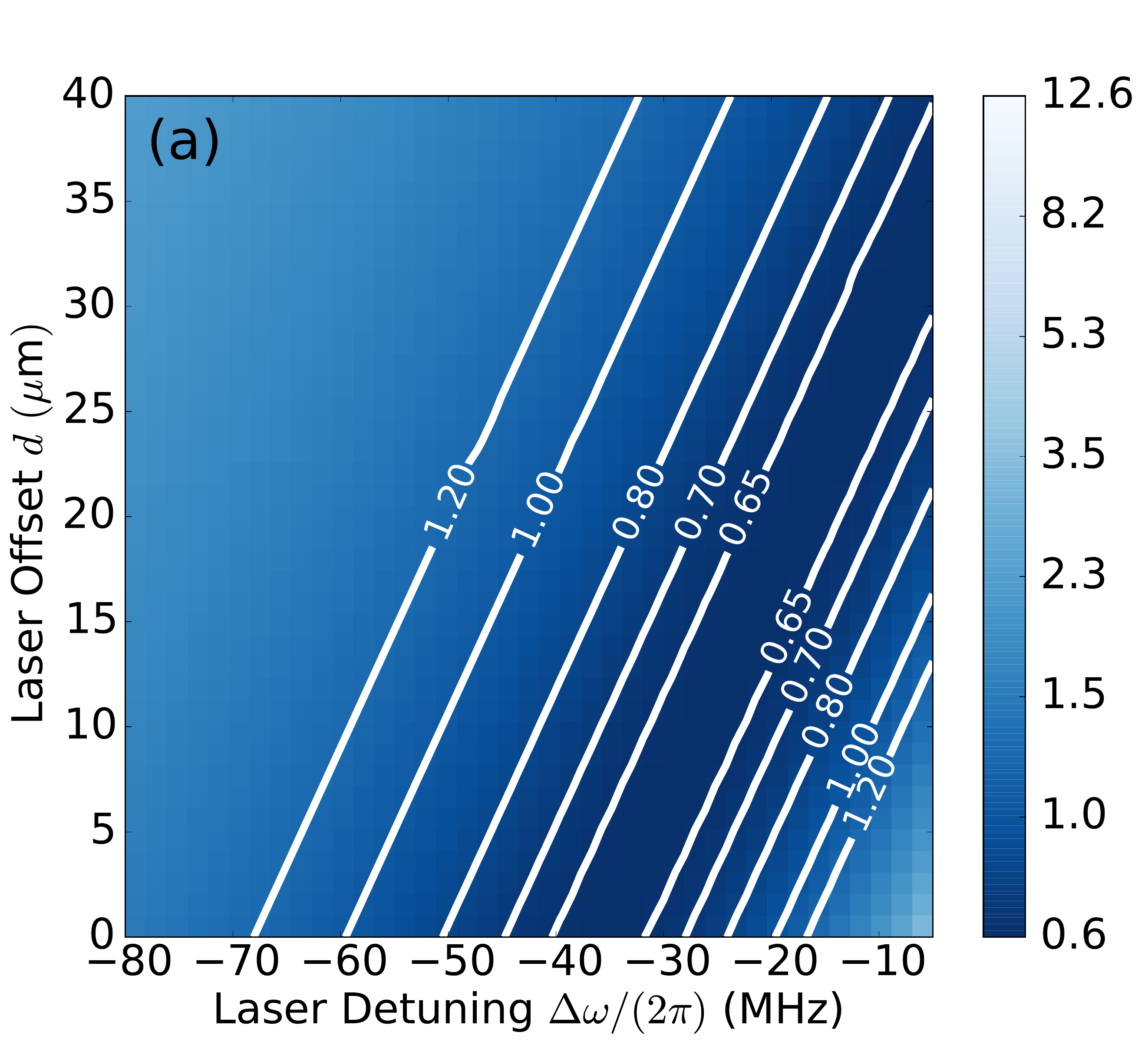}

\includegraphics[scale=0.49]{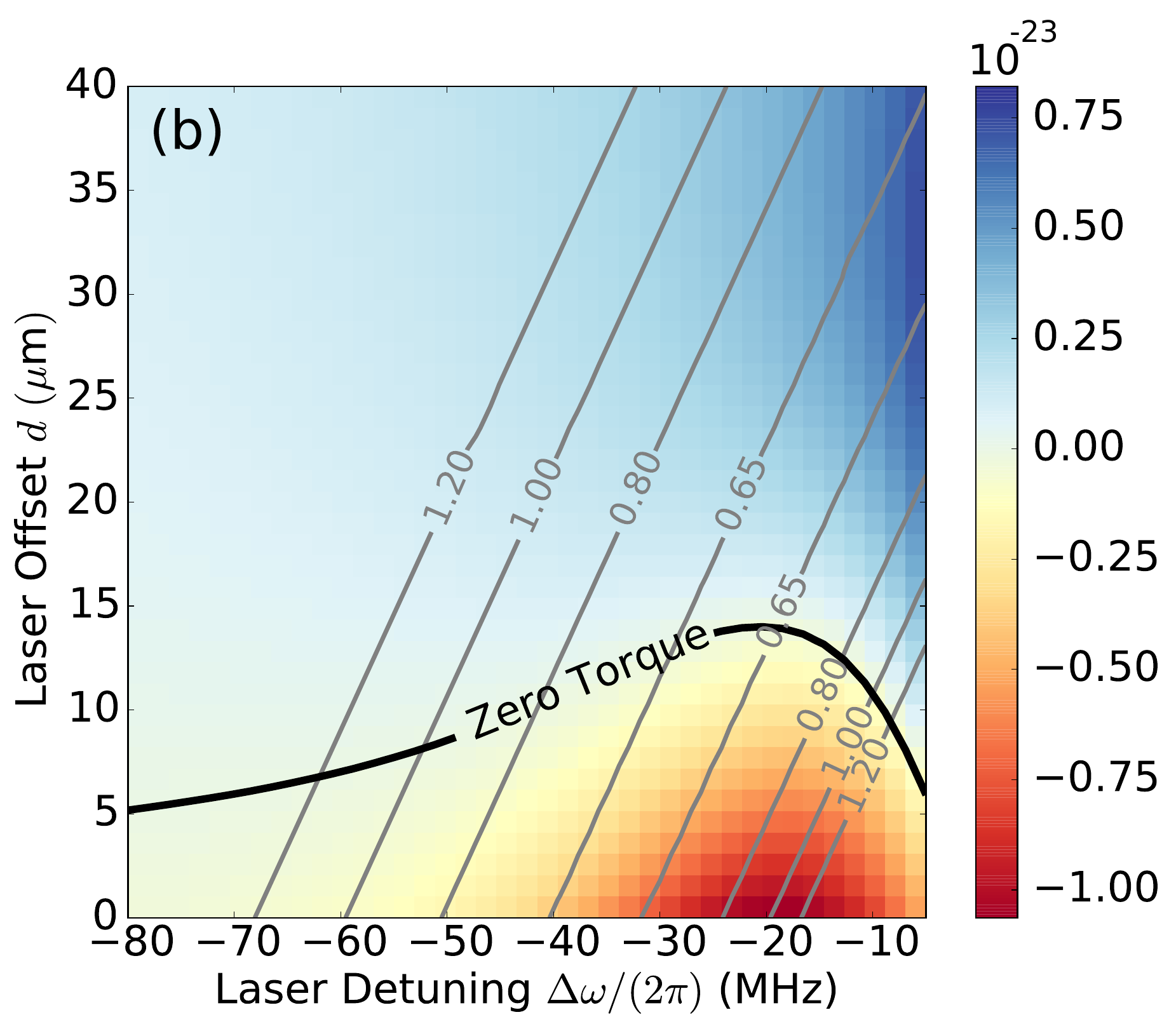}
\caption{(a) Contours of equilibrium planar temperature $T_{\perp}$ in millikelvin, plotted against laser detuning from the atomic transition frequency and offset $d$ of the laser beam from the center of the crystal. The calculation assumes $\omega_r/(2\pi) = 45$ kHz, $w_y = 30\, \mu\mathrm{m}$, $S_0=0.5$, $R_c=225\, \mu\mathrm{m}$, and values of $m$, and $\gamma_0/2$ appropriate for $^9$Be$^{+}$ ions. (b) Plot of the equilibrium net torque (in joules) imparted by the laser over the entire crystal. The same temperature contours seen in (a) are included. The torque calculation assumes $\Sigma_0 = 2.77 \times 10^9 \mathrm{m}^{-2}$.
\label{fig:30y45w}}
\end{figure}

Roots of Eq.\ (\ref{eqn:finalenergy}) provide Doppler laser cooling limits that do not require small laser waist $w_y$ and offset $d$. Figure \ref{fig:30y45w}(a) displays contours of the equilibrium planar temperature $T_{\perp}$ obtained from Eq.\ (\ref{eqn:finalenergy}), as a function of detuning $\Delta\omega$ of the laser beam from the atomic transition frequency and the offset $d$ of the laser beam from the center of the crystal. A rotation rate $\omega_r/(2\pi)= 45\:\mathrm{kHz}$, beam waist $w_y=30\:\mu\mathrm{m}$, and crystal radius $R_c= 225\:\mu\mathrm{m}$ is assumed.  For this plot $|d|, w_y \ll R_c$ is well satisfied, and we anticipate that the Doppler laser cooling limit should depend only on the combination $\Delta \omega -k \omega_r d$ (Eq.\ (\ref{eqn:detuning})). Indeed linear contour lines are observed, surrounding a low temperature ``trough'' where $T_{\perp}\sim 0.63 \: \mathrm{mK}$.  The observed slope of the contour lines is $1.12\:\mu\mathrm{m/MHz}$, in good agreement with the predicted slope of $\frac {2\pi} {k \omega_r} \left( 1 + \frac {w_y^2} {4R_c^2} \right) = 1.11\:\mu\mathrm{m/MHz}$. (Here we include the lowest order correction due to non-uniform density, given by Eq.\ (\ref{eqn:d prime}).) The figure documents that, with the rotating wall, low Doppler cooling limits can be obtained over a wide range of laser detunings and offsets.

Figure \ref{fig:30y45w}(b) shows the same temperature contour lines superimposed on a calculation of the torque imparted by the perpendicular cooling laser beam. This torque is balanced by a torque of equal magnitude but opposite sign generated by the rotating wall.  Because the rotating wall torque is applied on the boundary of the crystal while the laser torque is imparted in the crystal interior, conditions for zero laser torque should minimize shear stress, potentially important for producing stable crystals in which the ions do not move. Points of zero laser torque are indicated by the black curve in Fig.\ \ref{fig:30y45w}(b). Zero torque occurs for offsets $d>0$. This is because the finite waist $w_y$ of the laser beam means a small part of the beam interacts with ions on the $(y<0)$ side of the crystal.  Due to rotationally induced Doppler shifts, these ions scatter closer to resonance, leading to a balance with the torque generated by $(y>0)$ scattering. For the conditions of Fig.\ \ref{fig:30y45w}(b), the minimum $T_{\perp}\sim 0.63 \: \mathrm{mK}$ with zero laser torque is obtained for a laser detuning $\Delta \omega/(2\pi) \simeq 25\:\mathrm{MHz}$ ($\sim3$ half linewidths $\gamma_0 /4\pi$) and an offset $d \simeq 14\:\mu\mathrm{m}$ ($\sim0.5$ laser beam waist).

\begin{figure}
\includegraphics[scale=0.49]{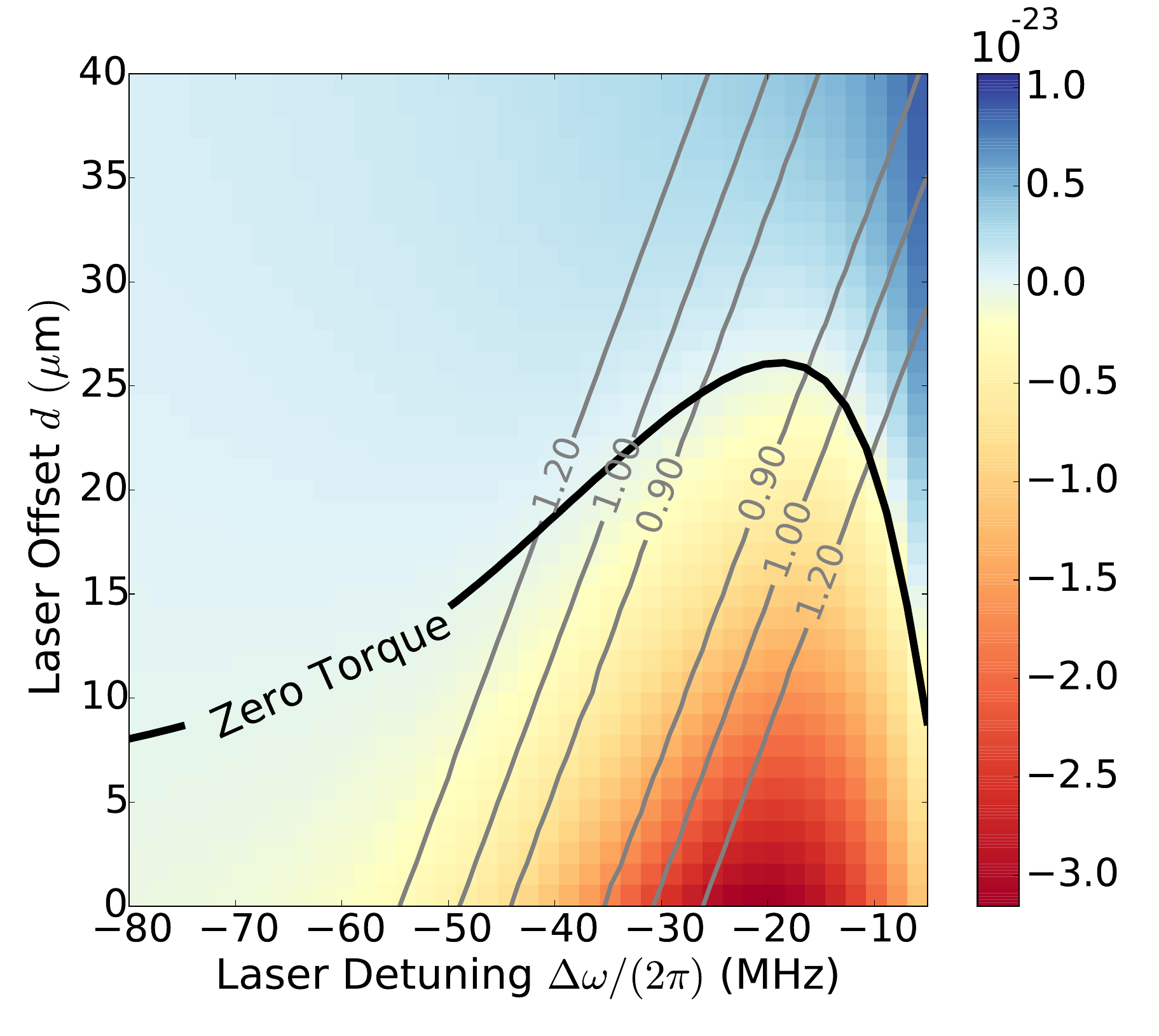}
\caption{Contours of planar temperature $T_{\perp}$ (in millikelvin) along with a color plot of the net torque (in joules) imparted by the perpendicular cooling laser. The calculation assumes $\omega_r/(2\pi) = 45$ kHz, $w_y = 60\, \mu\mathrm{m}$, $S_0=0.5$, $R_c=225\, \mu\mathrm{m}$, $\Sigma_0 = 2.77 \times 10^9 \mathrm{m}^{-2}$, and atomic parameters appropriate for $^9$Be$^{+}$ ions.
\label{fig:60y45w}}

\end{figure}

\begin{figure}
\includegraphics[scale=0.49]{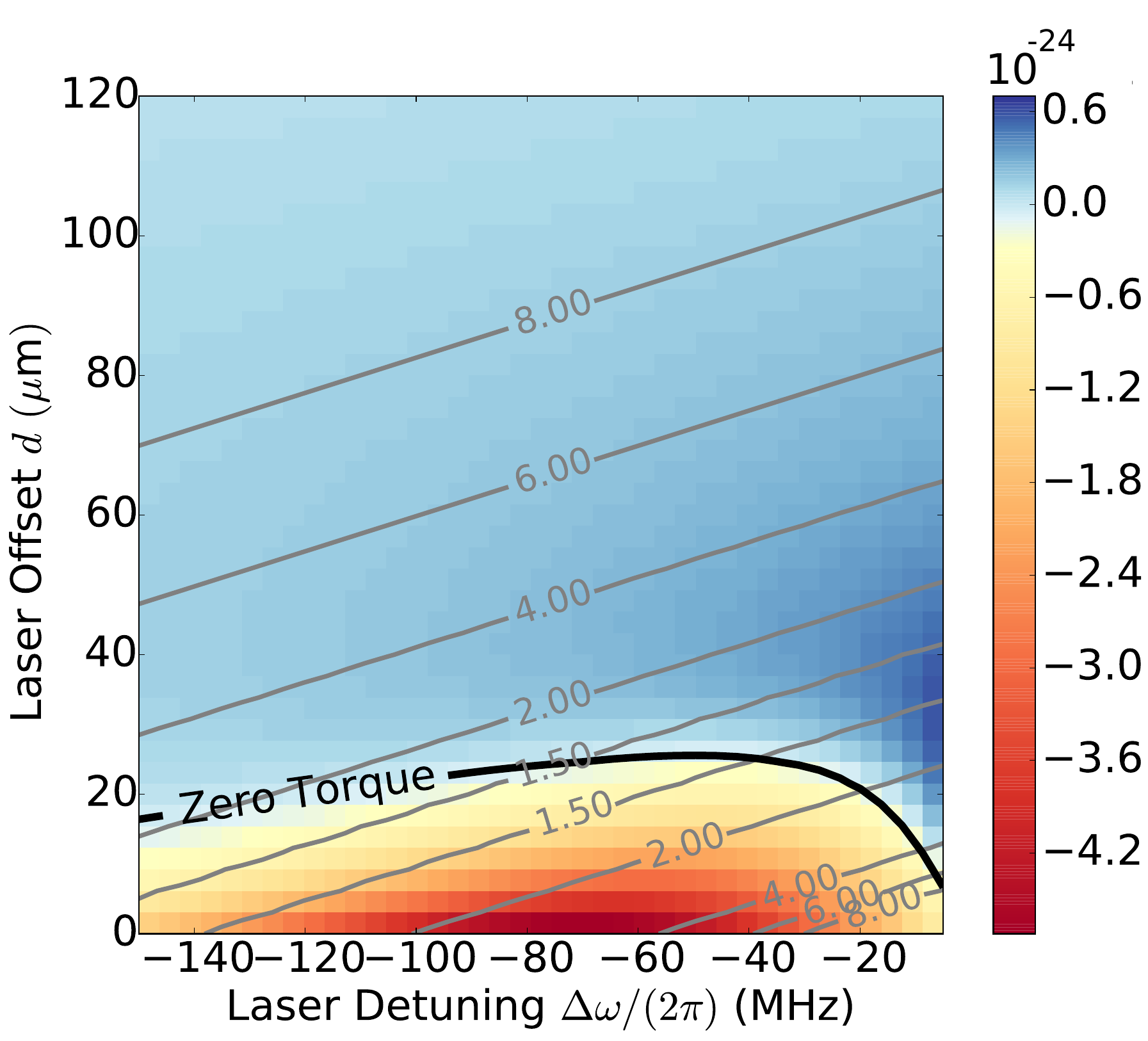}
\caption{Contours of planar temperature $T_{\perp}$ (in millikelvin) along with a color plot of the net torque (in joules) imparted by the perpendicular cooling laser. The calculation assumes $\omega_r/(2\pi) = 200$ kHz, $w_y = 30\, \mu\mathrm{m}$, $S_0=0.5$, $R_c=225\, \mu\mathrm{m}$, $\Sigma_0 = 2.77 \times 10^9 \mathrm{m}^{-2}$, and atomic parameters appropriate for $^9$Be$^{+}$ ions. We believe the small irregularities or waviness of the $1.50\: \mathrm{mK}$ contour lines are a numerical artifact.
\label{fig:200w}}

\end{figure}

Figures \ref{fig:60y45w} and \ref{fig:200w} show numerical calculations similar to Fig.\ \ref{fig:30y45w}(b) of the equilibrium temperature contours and the laser torque for a larger beam waist (Fig.\ \ref{fig:60y45w}) and a higher rotation frequency (Fig.\ \ref{fig:200w}).  Features similar to Figs.\ \ref{fig:30y45w}(a) and \ref{fig:30y45w}(b) are observed. In Fig.\ \ref{fig:60y45w}  the laser beam waist has been changed from $30\:\mu \mathrm{m}$ in Fig. \ref{fig:30y45w} to $60\:\mu \mathrm{m}$. The simplified  analysis of Eq.\ (\ref{eqn:dimensionless}) predicts temperature contours with the same slope as Fig.\ \ref{fig:30y45w}.  Linear and approximately parallel contour lines surrounding a low temperature trough where $T_{\perp}\sim 0.86 \: \mathrm{mK}$ are observed.  The slope of these contour lines is $1.38\:\mu\mathrm{m/MHz}$, which differs from the $1.12\:\mu\mathrm{m/MHz}$ contour line slope in Fig.\ \ref{fig:30y45w}, and also the predicted slope including the correction for the non-uniform density, $\frac {2\pi} {k \omega_r} \left( 1 + \frac {w_y^2} {4R_c^2} \right) = 1.13\:\mu\mathrm{m/MHz}$. This indicates that the $T_{\perp}$ obtained with a 60 $\mu\mathrm{m}$ laser beam waist ($\approx0.25 \times R_c)$ exhibits some deviations from the simplified analysis of Eqs.\ (\ref{eqn:dimensionless}) to (\ref{eqn:d prime}).  The minimum $T_{\perp}$ with zero laser torque is obtained for a laser detuning $\Delta \omega/(2\pi) \approx -21\:\mathrm{MHz}$ ($\sim2$ half linewidths $\gamma_0 /4\pi$) and an offset $d \approx 26\:\mu\mathrm{m}$ ($\sim0.5$ laser beam waist).

Recent NIST work uses an axial confinement (along the magnetic field) of $\sim 2$ MHz that results in rotation frequencies $\omega_r/(2\pi)\approx 200$ kHz \cite{bohj16}.  Figure \ref{fig:200w} shows the calculated $T_{\perp}$ for $\omega_y = 30 \: \mu \mathrm{m}$ (same as Fig. \ref{fig:30y45w}) but with $\omega_r/(2\pi) = 200$ kHz.  The measured slope of the contour lines, $0.25\:\mu\mathrm{m/MHz}$, agrees well with the predictions from the simplified analysis, $\frac {2\pi} {k \omega_r} \left( 1 + \frac {w_y^2} {4R_c^2} \right) = 0.25\:\mu\mathrm{m/MHz}$. The larger rotation frequency produces a larger dispersion $\Delta_{w}$ in the Doppler shift across the laser beam waist, resulting in an increase in the minimum temperature of $T_{\perp} \approx 1.45 \: \mathrm{mK}$.  The detuning that results in both the minimum temperature and the zero laser torque is significantly increased to $\Delta \omega/(2\pi) \approx -54\:\mathrm{MHz}$ ($\sim6$ half linewidths). The laser beam offset for this condition is slightly increased from Fig. \ref{fig:30y45w}, $d \approx 26\:\mu\mathrm{m}$ ($\sim 1$ laser beam waist). The greater sensitivity of $T_{\perp}$ to changes in the offset $d$ suggests that beam-pointing stability becomes a greater concern as the rotation frequency $\omega_r$ is increased.

\subsection{Recoil Heating with the Parallel Beam}

\begin{figure}[htb]
\includegraphics[scale=0.33]{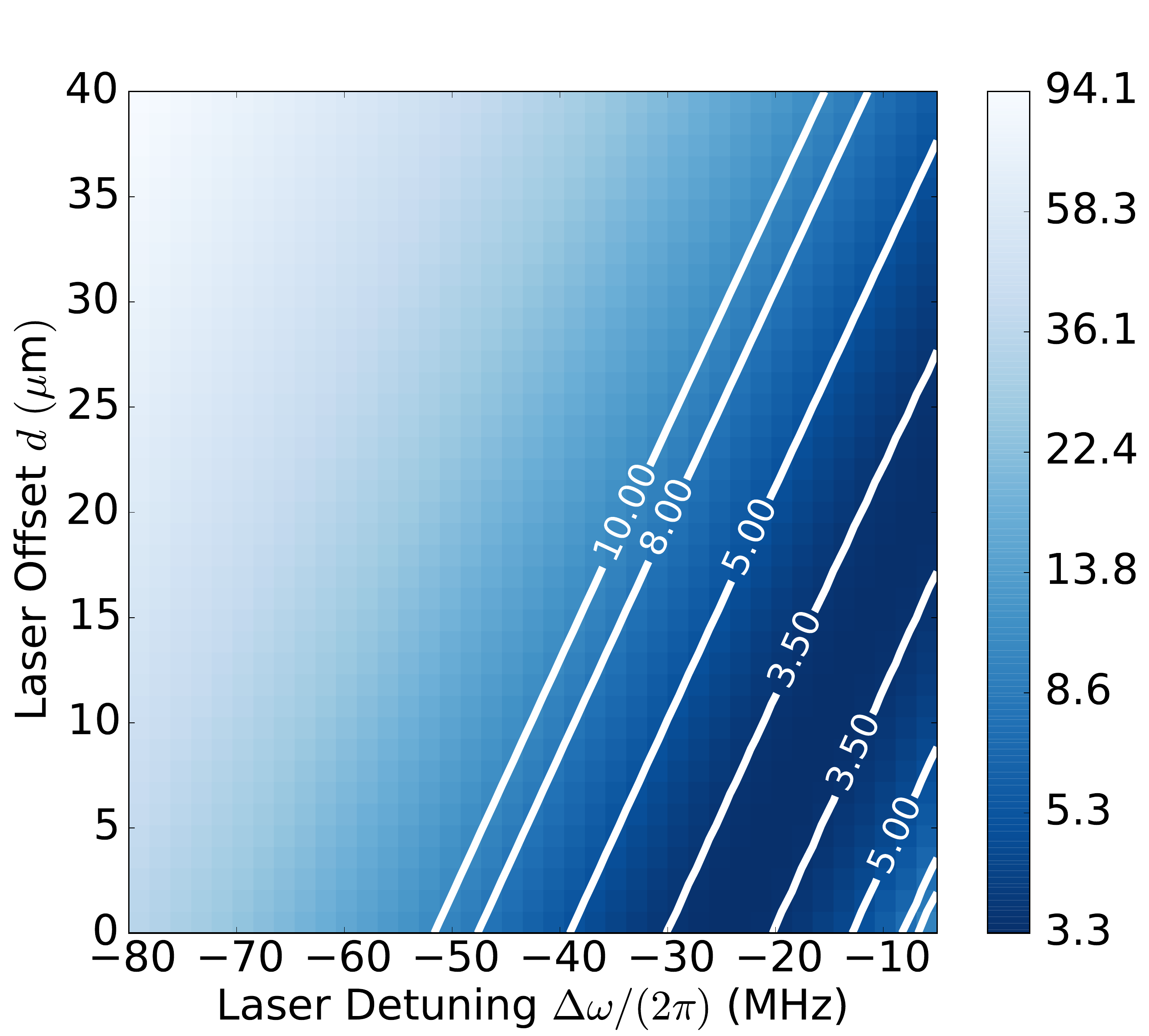}
\caption{In-plane equilibrium temperature $T_{\perp}$ (in millikelvin) for the identical parameters as Fig.\ \ref{fig:30y45w}, except for the addition of a parallel laser beam recoil heating term with $S_{\parallel}=0.2$. \label{fig:45wpar}}
\end{figure}

\begin{figure}[htb]
\includegraphics[scale=0.33]{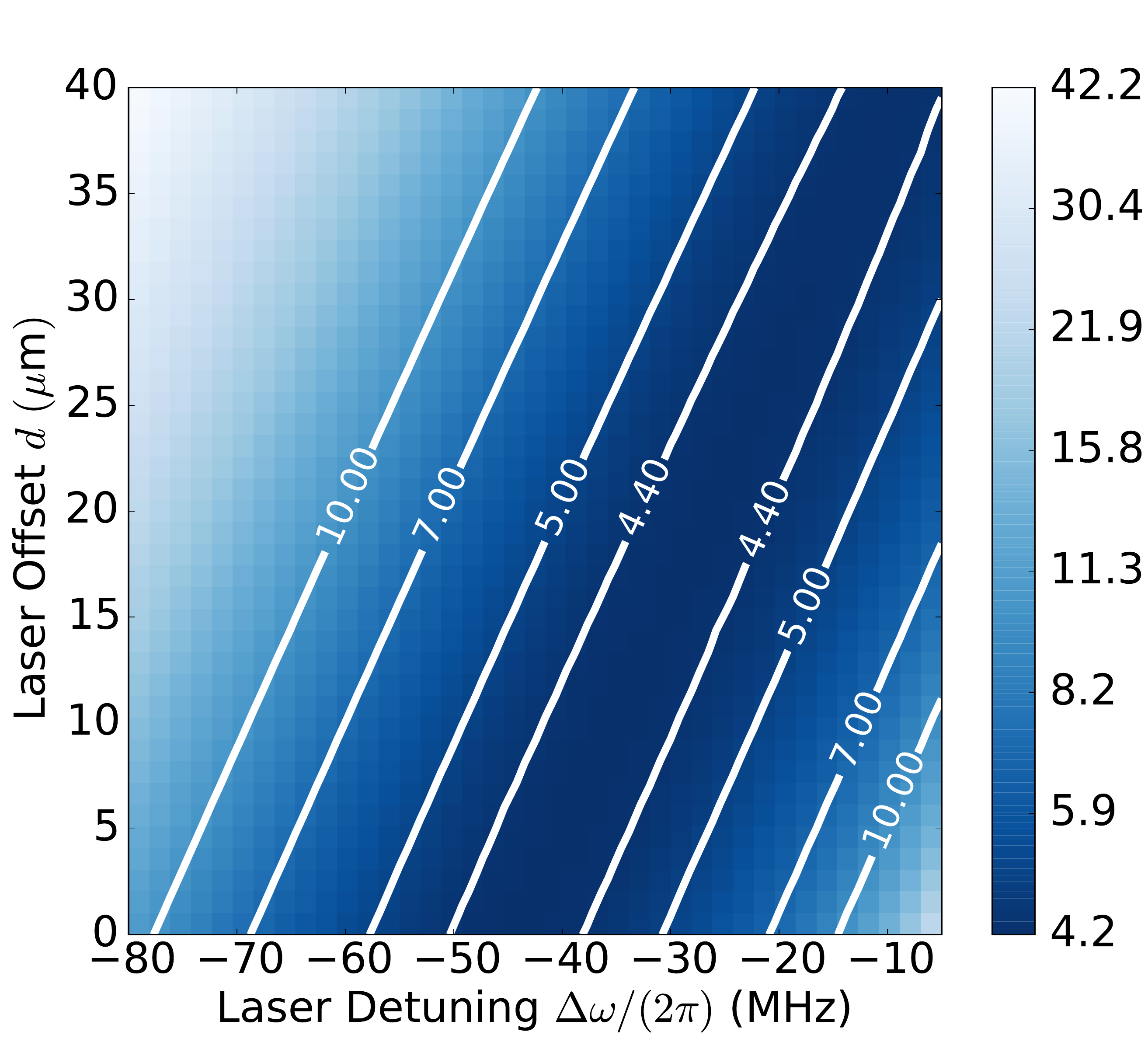}
\caption{In-plane equilibrium temperature $T_{\perp}$ (in millikelvin) for the identical parameters as Fig. \ref{fig:60y45w}, except for the addition of a parallel laser beam recoil heating term with $S_{\parallel}=0.2$. \label{fig:60wpar}}
\end{figure}

As discussed in Sec.\ \ref{sec:recoil}, experiments employ laser cooling with beams directed perpendicular and parallel to the magnetic field.  Doppler laser cooling of the parallel degrees of freedom is straightforward.  In contrast to the perpendicular motion, there is no coherent non-thermal motion and all ions scatter parallel laser light more or less equally.  This means the parallel laser beam scatter rate can be significantly larger than the perpendicular laser beam scatter rate.  Recoil heating of $T_{\perp}$ due to parallel laser light scattering is therefore a concern.

Figures \ref{fig:45wpar} and \ref{fig:60wpar} document that recoil heating from the parallel laser beam can be significant.  These figures show a calculation of $T_{\perp}$ for parameters identical to those used in Figs.\ \ref{fig:30y45w} and \ref{fig:60y45w}, but with recoil heating from a parallel laser beam with a saturation parameter of $S_{\parallel}=0.2$.  The basic features of a low-temperature trough surrounded by approximately linear temperature contours is maintained, but in both figures the minimum perpendicular temperature has increased by a factor of five.

\section{Conclusion \label{sec:conclusion}}
We updated existing theory for the laser cooling of 2-dimensional crystals of ions in a Penning trap to include energy exchange with a rotating wall potential. In contrast to theory and experimental measurements that did not use the rotating wall \cite{itaw88}, we find that low in-plane temperatures within a factor of two or three of the normal single-ion Doppler laser cooling limit can be obtained over a wide range of laser beam parameters and ion crystal rotation frequencies.

We also analyzed conditions for minimizing the torque imparted by the perpendicular laser-cooling beam. Low torque minimizes shear stress, which may increase the stability of the ion crystals. We find zero torque and the lowest $T_{\perp}$ are obtained with relatively large laser detunings $\Delta \omega$ that depend in detail on the ion crystal rotation frequency and laser beam waist. In general, optimization of the perpendicular laser beam cooling requires a different laser frequency and a set-up with greater flexibility than the parallel cooling. Finally, the relative strengths of the perpendicular and parallel laser beam scattering rates must be monitored to limit the impact of recoil heating from the parallel laser beam. This becomes particularly important as the crystal radius $R_c$ is increased, because for uniform parallel beam illumination, the recoil heating grows as $R_c^2$.

The theory discussed here neglected any coupling between the in-plane degrees of freedom and the transverse (or parallel to the $\hat{z}$-axis) degrees of freedom.  In addition it assumed that the in-plane degrees of freedom thermalize sufficiently rapidly that they can be described by thermal equilibrium characterized by a single temperature $T_{\perp}$. It would be useful to test the validity of these assumptions, which could be pursued through a first-principles simulation of the ion dynamics and the laser beam scattering. A first-principles simulation could also investigate the level of torque required to produce shear-induced instabilities of the ion positions within the crystal.

\section{Acknowledgements}
We acknowledge useful discussions with D. H. E. Dubin, R. C. Thompson, A. Keith, and D. Meiser.  We thank K. Gilmore and R. Fox for their comments on the manuscript. SBT was supported by Summer Undergraduate Research Fellowship funding through NIST. JGB was supported by a NIST NRC postdoctoral fellowship. This manuscript is a contribution of NIST and not subject to
U.S. copyright.

\bibliography{ion031316}{}

\begin{thebibliography}{26}
\expandafter\ifx\csname natexlab\endcsname\relax\def\natexlab#1{#1}\fi
\expandafter\ifx\csname bibnamefont\endcsname\relax
  \def\bibnamefont#1{#1}\fi
\expandafter\ifx\csname bibfnamefont\endcsname\relax
  \def\bibfnamefont#1{#1}\fi
\expandafter\ifx\csname citenamefont\endcsname\relax
  \def\citenamefont#1{#1}\fi
\expandafter\ifx\csname url\endcsname\relax
  \def\url#1{\texttt{#1}}\fi
\expandafter\ifx\csname urlprefix\endcsname\relax\def\urlprefix{URL }\fi
\providecommand{\bibinfo}[2]{#2}
\providecommand{\eprint}[2][]{\url{#2}}

\bibitem[{\citenamefont{Tan et~al.}(1995)\citenamefont{Tan, Bollinger,
  Jelenkovi\'{c}, and Wineland}}]{tanj95}
\bibinfo{author}{\bibfnamefont{J.~N.} \bibnamefont{Tan}},
  \bibinfo{author}{\bibfnamefont{J.~J.} \bibnamefont{Bollinger}},
  \bibinfo{author}{\bibfnamefont{B.}~\bibnamefont{Jelenkovi\'{c}}},
  \bibnamefont{and} \bibinfo{author}{\bibfnamefont{D.~J.}
  \bibnamefont{Wineland}}, \bibinfo{journal}{Phys. Rev. Lett.}
  \textbf{\bibinfo{volume}{75}}, \bibinfo{pages}{4198} (\bibinfo{year}{1995}).

\bibitem[{\citenamefont{Gruber et~al.}(2001)\citenamefont{Gruber, Holder,
  Steiger, Beck, DeWitt, Glassman, McDonald, Church, and Schneider}}]{grul01}
\bibinfo{author}{\bibfnamefont{L.}~\bibnamefont{Gruber}},
  \bibinfo{author}{\bibfnamefont{J.~P.} \bibnamefont{Holder}},
  \bibinfo{author}{\bibfnamefont{J.}~\bibnamefont{Steiger}},
  \bibinfo{author}{\bibfnamefont{B.~R.} \bibnamefont{Beck}},
  \bibinfo{author}{\bibfnamefont{H.~E.} \bibnamefont{DeWitt}},
  \bibinfo{author}{\bibfnamefont{J.}~\bibnamefont{Glassman}},
  \bibinfo{author}{\bibfnamefont{J.~W.} \bibnamefont{McDonald}},
  \bibinfo{author}{\bibfnamefont{D.~A.} \bibnamefont{Church}},
  \bibnamefont{and}
  \bibinfo{author}{\bibfnamefont{D.}~\bibnamefont{Schneider}},
  \bibinfo{journal}{Phys. Rev. Lett.} \textbf{\bibinfo{volume}{86}},
  \bibinfo{pages}{636} (\bibinfo{year}{2001}).

\bibitem[{\citenamefont{Shiga et~al.}(2011)\citenamefont{Shiga, Itano, and
  Bollinger}}]{shin11}
\bibinfo{author}{\bibfnamefont{N.}~\bibnamefont{Shiga}},
  \bibinfo{author}{\bibfnamefont{W.~M.} \bibnamefont{Itano}}, \bibnamefont{and}
  \bibinfo{author}{\bibfnamefont{J.~J.} \bibnamefont{Bollinger}},
  \bibinfo{journal}{Phys. Rev. A} \textbf{\bibinfo{volume}{84}},
  \bibinfo{pages}{012510} (\bibinfo{year}{2011}).

\bibitem[{\citenamefont{Andelkovic et~al.}(2013)\citenamefont{Andelkovic,
  Cazan, Nortershauser, Bharadia, Segal, Thompson, Johren, Vollbrecht, Hannen,
  and Vogel}}]{andz13}
\bibinfo{author}{\bibfnamefont{Z.}~\bibnamefont{Andelkovic}},
  \bibinfo{author}{\bibfnamefont{R.}~\bibnamefont{Cazan}},
  \bibinfo{author}{\bibfnamefont{W.}~\bibnamefont{Nortershauser}},
  \bibinfo{author}{\bibfnamefont{S.}~\bibnamefont{Bharadia}},
  \bibinfo{author}{\bibfnamefont{D.~M.} \bibnamefont{Segal}},
  \bibinfo{author}{\bibfnamefont{R.~C.} \bibnamefont{Thompson}},
  \bibinfo{author}{\bibfnamefont{R.}~\bibnamefont{Johren}},
  \bibinfo{author}{\bibfnamefont{J.}~\bibnamefont{Vollbrecht}},
  \bibinfo{author}{\bibfnamefont{V.}~\bibnamefont{Hannen}}, \bibnamefont{and}
  \bibinfo{author}{\bibfnamefont{M.}~\bibnamefont{Vogel}},
  \bibinfo{journal}{Phys. Rev. A} \textbf{\bibinfo{volume}{87}},
  \bibinfo{pages}{033423} (\bibinfo{year}{2013}).

\bibitem[{\citenamefont{Biercuk et~al.}(2009)\citenamefont{Biercuk, Uys,
  Devender, Shiga, Itano, and Bollinger}}]{biem09b}
\bibinfo{author}{\bibfnamefont{M.~J.} \bibnamefont{Biercuk}},
  \bibinfo{author}{\bibfnamefont{H.}~\bibnamefont{Uys}},
  \bibinfo{author}{\bibfnamefont{A.~P.~V.} \bibnamefont{Devender}},
  \bibinfo{author}{\bibfnamefont{N.}~\bibnamefont{Shiga}},
  \bibinfo{author}{\bibfnamefont{W.~M.} \bibnamefont{Itano}}, \bibnamefont{and}
  \bibinfo{author}{\bibfnamefont{J.~J.} \bibnamefont{Bollinger}},
  \bibinfo{journal}{Quan. Inf. Comp.} \textbf{\bibinfo{volume}{9}},
  \bibinfo{pages}{920} (\bibinfo{year}{2009}).

\bibitem[{\citenamefont{Britton et~al.}(2012)\citenamefont{Britton, Sawyer,
  Keith, Wang, Freericks, Uys, Biercuk, and Bollinger}}]{brij12}
\bibinfo{author}{\bibfnamefont{J.~W.} \bibnamefont{Britton}},
  \bibinfo{author}{\bibfnamefont{B.~C.} \bibnamefont{Sawyer}},
  \bibinfo{author}{\bibfnamefont{A.}~\bibnamefont{Keith}},
  \bibinfo{author}{\bibfnamefont{C.-C.~J.} \bibnamefont{Wang}},
  \bibinfo{author}{\bibfnamefont{J.~K.} \bibnamefont{Freericks}},
  \bibinfo{author}{\bibfnamefont{H.}~\bibnamefont{Uys}},
  \bibinfo{author}{\bibfnamefont{M.~J.} \bibnamefont{Biercuk}},
  \bibnamefont{and}
  \bibinfo{author}{\bibfnamefont{J.}~\bibnamefont{Bollinger}},
  \bibinfo{journal}{Nature} \textbf{\bibinfo{volume}{484}},
  \bibinfo{pages}{489} (\bibinfo{year}{2012}).

\bibitem[{\citenamefont{Wang et~al.}(2013)\citenamefont{Wang, Keith, and
  Freericks}}]{wanc13}
\bibinfo{author}{\bibfnamefont{C.-C.} \bibnamefont{Wang}},
  \bibinfo{author}{\bibfnamefont{A.~C.} \bibnamefont{Keith}}, \bibnamefont{and}
  \bibinfo{author}{\bibfnamefont{J.~K.} \bibnamefont{Freericks}},
  \bibinfo{journal}{Phys. Rev. A} \textbf{\bibinfo{volume}{87}},
  \bibinfo{pages}{013422} (\bibinfo{year}{2013}).

\bibitem[{\citenamefont{Mavadia et~al.}(2013)\citenamefont{Mavadia, Goodwin,
  Stutter, Bharadia, Crick, Segal, and Thompson}}]{mavs13}
\bibinfo{author}{\bibfnamefont{S.}~\bibnamefont{Mavadia}},
  \bibinfo{author}{\bibfnamefont{J.~F.} \bibnamefont{Goodwin}},
  \bibinfo{author}{\bibfnamefont{G.}~\bibnamefont{Stutter}},
  \bibinfo{author}{\bibfnamefont{S.}~\bibnamefont{Bharadia}},
  \bibinfo{author}{\bibfnamefont{D.~R.} \bibnamefont{Crick}},
  \bibinfo{author}{\bibfnamefont{D.~M.} \bibnamefont{Segal}}, \bibnamefont{and}
  \bibinfo{author}{\bibfnamefont{R.~C.} \bibnamefont{Thompson}},
  \bibinfo{journal}{Nature Comm.} \textbf{\bibinfo{volume}{4}},
  \bibinfo{pages}{2571} (\bibinfo{year}{2013}).

\bibitem[{\citenamefont{Bohnet et~al.}(2016)\citenamefont{Bohnet, Sawyer,
  Britton, Wall, Rey, Foss-Feig, and Bollinger}}]{bohj16}
\bibinfo{author}{\bibfnamefont{J.~G.} \bibnamefont{Bohnet}},
  \bibinfo{author}{\bibfnamefont{B.~C.} \bibnamefont{Sawyer}},
  \bibinfo{author}{\bibfnamefont{J.~W.} \bibnamefont{Britton}},
  \bibinfo{author}{\bibfnamefont{M.~L.} \bibnamefont{Wall}},
  \bibinfo{author}{\bibfnamefont{A.~M.} \bibnamefont{Rey}},
  \bibinfo{author}{\bibfnamefont{M.}~\bibnamefont{Foss-Feig}},
  \bibnamefont{and} \bibinfo{author}{\bibfnamefont{J.~J.}
  \bibnamefont{Bollinger}}, \bibinfo{journal}{arXiv:1512.03756}
  (\bibinfo{year}{2016}).

\bibitem[{\citenamefont{Anderegg et~al.}(2009)\citenamefont{Anderegg, Dubin,
  O'Neil, and Driscoll}}]{andf09}
\bibinfo{author}{\bibfnamefont{F.}~\bibnamefont{Anderegg}},
  \bibinfo{author}{\bibfnamefont{D.~H.~E.} \bibnamefont{Dubin}},
  \bibinfo{author}{\bibfnamefont{T.~M.} \bibnamefont{O'Neil}},
  \bibnamefont{and} \bibinfo{author}{\bibfnamefont{C.~F.}
  \bibnamefont{Driscoll}}, \bibinfo{journal}{Phys. Rev. Lett.}
  \textbf{\bibinfo{volume}{102}}, \bibinfo{pages}{185001}
  (\bibinfo{year}{2009}).

\bibitem[{\citenamefont{Jensen et~al.}(2005)\citenamefont{Jensen, Hasegawa,
  Bollinger, and Dubin}}]{jenm05}
\bibinfo{author}{\bibfnamefont{M.~J.} \bibnamefont{Jensen}},
  \bibinfo{author}{\bibfnamefont{T.}~\bibnamefont{Hasegawa}},
  \bibinfo{author}{\bibfnamefont{J.~J.} \bibnamefont{Bollinger}},
  \bibnamefont{and} \bibinfo{author}{\bibfnamefont{D.~H.~E.}
  \bibnamefont{Dubin}}, \bibinfo{journal}{Phys. Rev. Lett.}
  \textbf{\bibinfo{volume}{94}}, \bibinfo{pages}{025001}
  (\bibinfo{year}{2005}).

\bibitem[{\citenamefont{Sawyer et~al.}(2012)\citenamefont{Sawyer, Britton,
  Keith, Wang, Freericks, Uys, Biercuk, and Bollinger}}]{sawb12}
\bibinfo{author}{\bibfnamefont{B.~C.} \bibnamefont{Sawyer}},
  \bibinfo{author}{\bibfnamefont{J.~W.} \bibnamefont{Britton}},
  \bibinfo{author}{\bibfnamefont{A.~C.} \bibnamefont{Keith}},
  \bibinfo{author}{\bibfnamefont{C.-C.~J.} \bibnamefont{Wang}},
  \bibinfo{author}{\bibfnamefont{J.~K.} \bibnamefont{Freericks}},
  \bibinfo{author}{\bibfnamefont{H.}~\bibnamefont{Uys}},
  \bibinfo{author}{\bibfnamefont{M.~J.} \bibnamefont{Biercuk}},
  \bibnamefont{and} \bibinfo{author}{\bibfnamefont{J.~J.}
  \bibnamefont{Bollinger}}, \bibinfo{journal}{Phys. Rev. Lett.}
  \textbf{\bibinfo{volume}{108}}, \bibinfo{pages}{213003}
  (\bibinfo{year}{2012}).

\bibitem[{\citenamefont{Sawyer et~al.}(2014)\citenamefont{Sawyer, Britton, and
  Bollinger}}]{sawb14}
\bibinfo{author}{\bibfnamefont{B.~C.} \bibnamefont{Sawyer}},
  \bibinfo{author}{\bibfnamefont{J.~W.} \bibnamefont{Britton}},
  \bibnamefont{and} \bibinfo{author}{\bibfnamefont{J.~J.}
  \bibnamefont{Bollinger}}, \bibinfo{journal}{Phys. Rev. A}
  \textbf{\bibinfo{volume}{89}}, \bibinfo{pages}{033408}
  (\bibinfo{year}{2014}).

\bibitem[{\citenamefont{Itano and Wineland}(1982)}]{itaw82}
\bibinfo{author}{\bibfnamefont{W.~M.} \bibnamefont{Itano}} \bibnamefont{and}
  \bibinfo{author}{\bibfnamefont{D.~J.} \bibnamefont{Wineland}},
  \bibinfo{journal}{Phys. Rev. A} \textbf{\bibinfo{volume}{25}},
  \bibinfo{pages}{35} (\bibinfo{year}{1982}).

\bibitem[{\citenamefont{Itano et~al.}(1988)\citenamefont{Itano, Brewer, Larson,
  and Wineland}}]{itaw88}
\bibinfo{author}{\bibfnamefont{W.~M.} \bibnamefont{Itano}},
  \bibinfo{author}{\bibfnamefont{L.~R.} \bibnamefont{Brewer}},
  \bibinfo{author}{\bibfnamefont{D.~J.} \bibnamefont{Larson}},
  \bibnamefont{and} \bibinfo{author}{\bibfnamefont{D.~J.}
  \bibnamefont{Wineland}}, \bibinfo{journal}{Phys. Rev. A}
  \textbf{\bibinfo{volume}{38}}, \bibinfo{pages}{5698} (\bibinfo{year}{1988}).

\bibitem[{\citenamefont{Hendricks et~al.}(2008)\citenamefont{Hendricks,
  Phillips, Segal, and Thompson}}]{hen08}
\bibinfo{author}{\bibfnamefont{R.~J.} \bibnamefont{Hendricks}},
  \bibinfo{author}{\bibfnamefont{E.}~\bibnamefont{Phillips}},
  \bibinfo{author}{\bibfnamefont{D.~M.} \bibnamefont{Segal}}, \bibnamefont{and}
  \bibinfo{author}{\bibfnamefont{R.~C.} \bibnamefont{Thompson}},
  \bibinfo{journal}{J. Phys. B} \textbf{\bibinfo{volume}{41}},
  \bibinfo{pages}{035301} (\bibinfo{year}{2008}).

\bibitem[{\citenamefont{Asprusten et~al.}(2014)\citenamefont{Asprusten,
  Worthington, and Thompson}}]{aspm14}
\bibinfo{author}{\bibfnamefont{M.}~\bibnamefont{Asprusten}},
  \bibinfo{author}{\bibfnamefont{S.}~\bibnamefont{Worthington}},
  \bibnamefont{and} \bibinfo{author}{\bibfnamefont{R.~C.}
  \bibnamefont{Thompson}}, \bibinfo{journal}{Appl. Phys. B}
  \textbf{\bibinfo{volume}{114}}, \bibinfo{pages}{157} (\bibinfo{year}{2014}).

\bibitem[{\citenamefont{Huang et~al.}(1997)\citenamefont{Huang, Anderegg,
  Hollmann, Driscoll, and O'Neil}}]{huap97}
\bibinfo{author}{\bibfnamefont{X.-P.} \bibnamefont{Huang}},
  \bibinfo{author}{\bibfnamefont{F.}~\bibnamefont{Anderegg}},
  \bibinfo{author}{\bibfnamefont{E.~M.} \bibnamefont{Hollmann}},
  \bibinfo{author}{\bibfnamefont{C.~F.} \bibnamefont{Driscoll}},
  \bibnamefont{and} \bibinfo{author}{\bibfnamefont{T.~M.}
  \bibnamefont{O'Neil}}, \bibinfo{journal}{Phys. Rev. Lett.}
  \textbf{\bibinfo{volume}{78}}, \bibinfo{pages}{875} (\bibinfo{year}{1997}).

\bibitem[{\citenamefont{Huang et~al.}(1998)\citenamefont{Huang, Bollinger,
  Mitchell, Itano, and Dubin}}]{huap98b}
\bibinfo{author}{\bibfnamefont{X.-P.} \bibnamefont{Huang}},
  \bibinfo{author}{\bibfnamefont{J.~J.} \bibnamefont{Bollinger}},
  \bibinfo{author}{\bibfnamefont{T.~B.} \bibnamefont{Mitchell}},
  \bibinfo{author}{\bibfnamefont{W.~M.} \bibnamefont{Itano}}, \bibnamefont{and}
  \bibinfo{author}{\bibfnamefont{D.~H.~E.} \bibnamefont{Dubin}},
  \bibinfo{journal}{Phys. Plasmas} \textbf{\bibinfo{volume}{5}},
  \bibinfo{pages}{1656} (\bibinfo{year}{1998}).

\bibitem[{\citenamefont{Mitchell et~al.}(1998)\citenamefont{Mitchell,
  Bollinger, Dubin, Huang, Itano, and Baughman}}]{mitt98}
\bibinfo{author}{\bibfnamefont{T.~B.} \bibnamefont{Mitchell}},
  \bibinfo{author}{\bibfnamefont{J.~J.} \bibnamefont{Bollinger}},
  \bibinfo{author}{\bibfnamefont{D.~H.~E.} \bibnamefont{Dubin}},
  \bibinfo{author}{\bibfnamefont{X.-P.} \bibnamefont{Huang}},
  \bibinfo{author}{\bibfnamefont{W.~M.} \bibnamefont{Itano}}, \bibnamefont{and}
  \bibinfo{author}{\bibfnamefont{R.~H.} \bibnamefont{Baughman}},
  \bibinfo{journal}{Science} \textbf{\bibinfo{volume}{282}},
  \bibinfo{pages}{1290} (\bibinfo{year}{1998}).

\bibitem[{\citenamefont{Bharadia et~al.}(2012)\citenamefont{Bharadia, Vogel,
  Segal, and Thompson}}]{bhas12}
\bibinfo{author}{\bibfnamefont{S.}~\bibnamefont{Bharadia}},
  \bibinfo{author}{\bibfnamefont{M.}~\bibnamefont{Vogel}},
  \bibinfo{author}{\bibfnamefont{D.~M.} \bibnamefont{Segal}}, \bibnamefont{and}
  \bibinfo{author}{\bibfnamefont{R.~C.} \bibnamefont{Thompson}},
  \bibinfo{journal}{Appl. Phys. B} \textbf{\bibinfo{volume}{107}},
  \bibinfo{pages}{1105} (\bibinfo{year}{2012}).

\bibitem[{\citenamefont{Dubin and O'Neil}(1999)}]{dubd99}
\bibinfo{author}{\bibfnamefont{D.~H.~E.} \bibnamefont{Dubin}} \bibnamefont{and}
  \bibinfo{author}{\bibfnamefont{T.~M.} \bibnamefont{O'Neil}},
  \bibinfo{journal}{Rev. Mod. Phys.} \textbf{\bibinfo{volume}{71}},
  \bibinfo{pages}{87} (\bibinfo{year}{1999}).

\bibitem[{foo()}]{footnote.thermalCOM}
\bibinfo{note}{We note that for a quadratic trap potential thermal
  equilibration of the magnetron and cyclotron center-of-mass modes may not be
  a reasonable assumption.}

\bibitem[{\citenamefont{Dubin}(2013)}]{dubd13}
\bibinfo{author}{\bibfnamefont{D.~H.~E.} \bibnamefont{Dubin}},
  \bibinfo{journal}{Phys. Rev. A} \textbf{\bibinfo{volume}{88}},
  \bibinfo{pages}{013403} (\bibinfo{year}{2013}).

\bibitem[{\citenamefont{Jensen et~al.}(2004)\citenamefont{Jensen, Hasegawa, and
  Bollinger}}]{jenm04}
\bibinfo{author}{\bibfnamefont{M.~J.} \bibnamefont{Jensen}},
  \bibinfo{author}{\bibfnamefont{T.}~\bibnamefont{Hasegawa}}, \bibnamefont{and}
  \bibinfo{author}{\bibfnamefont{J.~J.} \bibnamefont{Bollinger}},
  \bibinfo{journal}{Phys. Rev. A} \textbf{\bibinfo{volume}{70}},
  \bibinfo{pages}{033401} (\bibinfo{year}{2004}).

\bibitem[{\citenamefont{Mavadia et~al.}(2014)\citenamefont{Mavadia, Stutter,
  Goodwin, Crick, Thompson, and Segal}}]{mavs14}
\bibinfo{author}{\bibfnamefont{S.}~\bibnamefont{Mavadia}},
  \bibinfo{author}{\bibfnamefont{G.}~\bibnamefont{Stutter}},
  \bibinfo{author}{\bibfnamefont{J.~F.} \bibnamefont{Goodwin}},
  \bibinfo{author}{\bibfnamefont{D.~R.} \bibnamefont{Crick}},
  \bibinfo{author}{\bibfnamefont{R.~C.} \bibnamefont{Thompson}},
  \bibnamefont{and} \bibinfo{author}{\bibfnamefont{D.~M.} \bibnamefont{Segal}},
  \bibinfo{journal}{Phys. Rev. A} \textbf{\bibinfo{volume}{89}},
  \bibinfo{pages}{032502} (\bibinfo{year}{2014}).

\end{thebibliography}

\end{document}